# Ethical Implications of IT-enabled Information Flows Conceived as Intermediaries or Mediators

**Dubravka Cecez-Kecmanovic**
School of Information Systems, Technology and Management
University of NSW Business School, Sydney, Australia
Email: dubravka@unsw.edu.au

**Olivera Marjanovic**
Business Information Systems Discipline
University of Sydney Business School, Sydney, Australia
Email: olivera.marjanovic@sydney.edu.au

## Abstract

This paper contributes to a better understanding of ethical concerns regarding the deployment of complex public sector IT systems and the information flows they instigate. The paper aims to reveal how different views on IT and IT-enabled information flows allow us to see differently their social implications and to construe different ethical questions. This is achieved by i) defining two opposing views on IT-enabled information flows as 'intermediaries' and 'mediators'; ii) by analysing the controversial case of My School – a web portal that provides performance data of 9,500 Australian schools – that introduces new information flows in the education sector; and iii) by revealing and explaining how some unintended negative social implications emerge and how the articulation of ethical concerns depends on the view on My School-enabled information flows. The paper concludes with theoretical and practical implications, with particular emphasis on responsibilities of all involved, setting up foundations for an important area of future IS research.

## 1 Introduction

The deployment of Information Technology (IT) systems in the public sector – health care, education, social services, and others – is radically transforming information flows between government agencies, public sector organizations and citizens (Rodinalli 2007; Fox 2010; Keevers et al. 2012). While the logic of IT deployment in the public sector has largely followed the private sector's pursuit of efficiency and effectiveness (Mosse and Whitley 2009), it is accountability and transparency that have been the key objectives specific to the public sector (Keevers et al. 2012). However, while fulfilling these objectives IT systems also introduce new and emerging information flows between governments, public sector organizations and citizens which are reported to produce unintended negative consequences for citizens (and segments of community) (Kappos et al. 2005; Overby et al. 2010; Fichman et al. 2015; Tarafdar et al. 2015a, 2015b). This raises serious ethical questions that remain hidden in the rhetoric of efficiency, accountability and transparency.

This paper draws attention to and examines the ethics of IT systems in the public sector by focusing on the emerging information flows and their broader implications for "the good life within one's community" (Mingers and Walsham 2010, p. 841). When the implications are in any way harmful or negatively affect wellbeing of a community and its members the goodness or virtue of the IT system and its information flows have to be questioned and addressed (Mason 1995; Mingers and Walsham 2010). The problem however arises when these negative implications are unrecognized or disregarded (considered neglectable vis-à-vis necessity of digitization and clear evidence of positive effects), thus masking serious ethical issues and preventing necessary and due responses by all concerned. Central to this debate is the ethical question of public sector IT systems that reconfigure information flows between government, public sector organizations and citizens in complex, opaque and unpredictable ways without due understanding or appreciation of the emerging ethical consequences for a community.

Within the vast domain of ethical implications of IT systems in the public sector this paper focuses more specifically on the (im)possibility of identifying and articulating ethical concerns about IT-enabled information flows. The paper aims to reveal how different views on IT and IT-enabled information flows emerging in a community allow us to see differently their social implications and to construe different ethical questions. This is achieved by i) defining two opposing views on IT-enabled information flows conceptualized as 'intermediaries' or 'mediators'; ii) by analysing the case of My School, a web portal that provides performance data about 9,500 Australian schools that introduces new information flows between Government, schools, parents, students and other actors in the education sector; and iii) by revealing and explaining how some unintended negative social implications emerge in the education sector – for schools, students, parents, teachers, education institutions and the broader public – and how the articulation of ethical concerns differs depending on the view on My School-enabled information flows as intermediaries or mediators. The paper thus contributes to better understandings of ethical issues involved in IT-enabled information flows in a public sector and the critical role of conceptual views on these information flows in articulating and assessing their goodness and virtue for citizens and the broader community. The paper concludes with theoretical and practical implications, with particular emphasis on the responsibilities of all involved – Government agency, media, schools, education institutions, teachers, students and parents.

## 2 Background

Public sector IT systems in health care, education, social services and other public sector domains, are developed, implemented, regulated and managed by government or public sector agencies to serve the broader public and provide open data and services to citizens (Smith 1995; Denzinger et al. 2002; West 2004; Dunleavy et al. 2006). Often considered as a synonym of public sector modernization, the IT systems are deployed with the objectives of economic efficiency, accountability and transparency and also the provision of equitable services and benefits to citizens and the broader community. A distinct feature of the public sector IT systems is the introduction of new IT-enabled information flows between governments, public sector organizations and citizens as well as other actors in a society. By making relevant data (e.g. in health care or education) publicly available and allowing free access and unrestricted usage of data, these IT systems radically reconfigure information flows in a particular public sector in complex and unpredictable ways (see e.g. Smith 1995; Blum 2014; Salzberg 2014; Weaver et al. 2014). While economic efficiency, accountability and transparency typically justify the introduction of these new information flows, their unintended and often negative social implications for certain segments of society remain either unnoticed or neglected (Visscher and Coe 2003; Earl and Katz 2006; van der Hoven and Wickert 2008; Vanderlinde et al. 2010). In particular, the increasing implementation of IT-enabled information flows in the public sector has also introduced new inequalities and produced unfair and unjust (while unintended) implications for some sections of a community. These are among critical ethical questions that the Information Systems discipline has only begun to consider as relevant research problems (e.g. Stahl 2008; Davison et al. 2009; Mingers and Walsham 2010).

IS research has identified and began to address several important aspects of unexpected and unpredictable negative societal consequences of complex IT applications (Tarafdar et al. 2015a, pp.165). Importantly, broader social implications and significant negative unintended consequences of IT applications in the public sector create new ethical and moral challenges for modern democratic societies (Smith 1995; Mingers and Walsham 2010). More generally, IS researchers have drawn from ethical and moral theories to clarify and explain social implications of modern IT systems (e.g. Smith and Hasnas 1999; Stahl 2008; Davison et al. 2009; Mingers and Walsham 2010; Ross and Chiasson 2011). For instance well known ethical theories, such as *consequentialism* (Benthem 1948/1789; Mill 2002/1861), *deontology* (Kant 1991/1785; Rowls 1971; McNaughton and Rawling 2007), and *virtue ethics and communitarianism* (MacIntyre 1985; Hursthouse 2007) have been applied to examine ethical implication of IT systems and business transformation (Mingers and Walsham 2010). Contrary to such approaches, Mingers and Walsham (2010) apply Habermas' discourse ethics (1992), as a procedural approach to deal with ethical questions of implementing IT systems in a democratic community.

Drawing from the literature on social and ethical implications of complex IT systems in the community and also the related literature on theories of ethics, we conclude that a key issue remaining unexplored is the possibility, or impossibility, of identifying and articulating ethical concerns before we can examine ways of addressing them (e.g. apply an ethical theory). We propose that different conceptual views on IT systems in their social contexts allow us to see different social implications and thus articulate different ethical concerns. In this context we find particularly relevant the debate about conceptualizing things and technologies as *intermediaries* or *mediators* in Science and Technology Studies (Callon 1991; Latour 1992). Traditionally things and technologies (non-humans) were conceived as intermediaries, that is, black boxes that merely transport action from elsewhere; they were seen not as actors themselves, but as entities that stand in for real, actual actors. Callon (1991) critiqued this conception and argued that technologies and things were not mere intermediaries but more accurately mediators. Related with human actors, non-human actors modify relations between them. Latour (1996) further argued that "objects are not means, but rather mediators – just as other actors are. They do not transmit our force faithfully, any more than we are faithful messengers of theirs" (p. 240). Both Callon (1986, 1991) and Latour (1992, 2005) emphasised the importance of the distinction between intermediaries and mediators when explaining the role of technologies and things as conditions for the possibility of human societies.

We find this distinction between intermediaries and mediators particularly relevant in exploring IT-enabled information flows and their social and ethical implications. For instance, implicit in the deployment of IT systems in the public sector has been the assumption that IT systems are neutral tools or instruments (the means) to achieve highly desirable ends (Introna 2007). As a consequence IT-enabled information flows are seen as intermediaries that transmit information (meanings) from sources (IT systems) to users. As such information flows cause certain ways of doing things in practice and thus create more or less determinable and predictable impacts on the users. This further implies that if values, interests and purposes invested in an IT system are justified as fair and ethical, the impacts of IT-enabled information flows on users are also fair and ethical. Hence the task of ethics is to engage in disclosing and scrutinizing assumptions, interests, values and purposes built into IT-systems and transmitted to users. This is summarized in Table 1.

|  | **Tool or instrument view on IT systems** | **Performative view on IT systems** |
|---|---|---|
| **IT-enabled information flows** | Information flows are assumed to transmit information (meanings) from IT systems to its users and are thus seen as *intermediaries*; <br><br> Information flows cause changes in practices and therefore produce determinate and predictable impacts on users. | Information flows are *mediators* as they translate, reconstruct and distort information they supposedly transmit; <br><br> Embedded in users' practices information flows transform and are transformed by these practices, thus producing emerging and unpredictable social changes. |
| **Approach to ethical implications of information flows** | Ethical concerns are focused on IT systems and built-in assumptions, values, interest and purposes as well as the expected impacts of information flows on users; <br><br> The task of ethics is to disclose and scrutinize these assumptions, interests, values and purposes as well as assess the determinate effects of IT-enabled information flows on users. | Ethical concerns include the ongoing performative co-construction of information flows and practices and the ensuing framing of users and their identities; <br><br> The task of ethics is ontological disclosure: revealing and opening to scrutiny reconfigurations of users' practices and their identities and production of new social order. |

*Table 1 Conceptual framework that clarifies how the views on IT-enabled information flows as intermediaries vs mediators allow for different approaches to revealing and articulating their ethical implication*

An alternative performative view on IT systems assumes that technology and society co-construct each other (Cecez-Kecmanovic et al. 2014). When implemented and used, IT systems are performed in users' practices while simultaneously transforming both users and their practices. Consequently IT-enabled information flows are mediators as they translate, reconstruct, and distort information they supposedly transmit. Embedded in users' practices IT-enabled information flows reconfigure these practices, together with the users and their identities and as a result these information flows themselves change. The impacts of IT-enabled information flows are thus neither predictable nor determinable. The view on IT-enabled information flows as mediators radically changes the approach to ethical implications by broadening the perspective to the field of practice and the complex ways information flows and practice co-produce each other.

We can therefore propose that the task of ethics, in this context, is to reveal and open to scrutiny ongoing performance of IT-enabled information flows and the emerging reconfiguration and co-construction of users and their practices as well as the production of new a social order more generally (Introna 2002). In other words, the task of ethics is *ontological disclosure* (Introna 2007) through which these reconfigurations and reconstructions are uncovered and subjected to scrutiny.

The conceptual framework presented in Table 1 summarizes our proposition that different conceptual views on IT systems in their social contexts (and in particular IT-enabled information flows) allow us to see different social implications and thereby approach and articulate ethical implications differently. This conceptual framework allows us to frame new research questions that require empirical investigation: How and why do different social and ethical consequences show up when considering the effects of IT-enabled information flows in a social domain? What are the implications of these differences for all concerned? In the rest of the paper we seek to answer these questions by drawing on a case study that we describe next.

## 3    Methodology: An Interpretive Case Study of *My School*

We examine our research questions based on an interpretive case study of a web-based portal My School (www.myschool.edu.au) designed to provide school performance data based on the *National Assessment Program – Literacy and Numeracy* (NAPLAN) test. NAPLAN is administered at the same time in all Australian schools (currently over 9500) to Year 3, 5, 7 and 9 students. The Australian Curriculum Assessment Report Authority (ACARA) is a Government agency that planned, designed and introduced My School portal in 2010 to make school performance data (NAPLAN test results) publicly available. The objective was to achieve transparency and accountability of schools and teachers.

When launched in January 2010, My School included two sets of NAPLAN results for 2008 and 2009. Currently in its sixth year of operation the portal provides 8 years worth of data. In addition to data, My School also provides easy-to-use tools enabling any user (non-registered) to search and compare various aspects of schools' and students' performance over time. The outcomes of these operations are shown in simple visual forms to facilitate better understanding. For example, users could see and re-order various lists (by clicking on the column titles), look at graphical comparisons (e.g. showing the whole school in relation to the national average), or geographical maps.

My School is a highly relevant case for exploring our research questions. It is a government IT system that introduced new information flows in the education sector – an important and sensitive domain of social practice and policy. These new and emerging information flows are reconfiguring relations among the Government, schools and citizens with numerous intended and unintended consequences, creating unprecedented public controversy and criticism. For example, due to its unexpected and, in many documented cases, harmful effects for the intended beneficiaries (children, parents, teachers, and schools), My School was subjected to two Senate enquiries.

Another particularly attractive feature of the My School case is the public availability of high quality and rich data sets from very reputable public sources maintained and provided by different government agencies. This is because My School portal is owned and managed by a government agency, and as such is regulated and closely monitored.

Data collection for this study spans a period of over nine years (2006 – 2015), starting well before the official launch of My School portal in 2010 and continuing until present day. Therefore, our data set covers the initial planning of a national numeracy and literacy test (NAPLAN test), pre-My School implementation of NAPLAN testing over two years (2008, 2009), leading to My School development and its launch in 2010, and the ongoing use of this web portal. In addition to Government data, we collected articles and public debates in media, public responses and video cases posted on various web sites (Government, school principles' association, teachers' associations), blogs and twitter feeds as well as numerous published studies completed by researchers in other disciplines (such as public policy, social science, politics, education and so on). Our resulting data set consists of 400+ documents and is growing by the day. Table 2 (in the appendix) includes a selected sample of the most relevant documents and key My School-related events (chronologically ordered), which are used in the study and presented in this paper.

Our research methodology is interpretive, informed by hermeneutics as both a philosophy and a methodology for analysing texts and interpreting actions and meanings (Gadamer 1960; Crotty 1998). Our interpretation emerged gradually, through a dialogical engagement with collected evidence and observation of actions by the growing number of actors (parents, children, teachers, school principles, government agencies, media, financial experts, researchers, politicians etc.). This interpretative process was iterative in a sense that understanding was constantly moving from the whole to the part and back to the whole – referred to as a hermeneutic circle (Gadamer 1960; Klein and Myers 1999). As such our interpretation remained provisional and progressive, never to be finally correct as 'the harmony of all the details' (Gadamer 1960) can never be reached. Rather through the hermeneutic way of understanding we sought to create a rich picture of different views on information flows in the education sector. In addition to the view expressed by ACARA and explicitly articulated in the My School portal we identified the views expressed by numerous other actors (users) in the education sectors. Importantly, our hermeneutic analysis was informed by the conceptual framework described above and guided by our research questions.

Data analysis was conducted through a number of hermeneutic circles. We read the documents as we collected them and classified them according to the source, authority and medium, document purpose and related event, and the topics addressed. For this paper we selected documents related to key events (as listed in Table 2 in the Appendix) and in particular identified those related to information flows. Within the selected documents, we coded sections of the text that refer to some important aspects of My School-enabled information flows, including for instance, how schools became labelled "good" or "bad" schools after school league table were published by newspapers; how school practices changed and how students and parents experienced school responses to students' NAPLAN test results (revealed in testimonies during Senate enquires). We compared these views with those expressed on My School web portal and also by ACARA officials in media. This analysis allowed us to explore numerous consequences of My School use, both intended and unintended, and answer our research questions.

## 4 Findings

This story began in July 2006 when the Ministerial Council on Education, Employment, Training and Youth Affairs (MCEETYA) formally approved the introduction of nation-wide testing in Literacy and Numeracy to all Australian schools. This led to the development of the NAPLAN test during 2007, with the main objective "to provide a measure of: (a) how individual students are performing at the time of the tests; (b) whether or not young people are meeting literacy and numeracy benchmarks, and (c) how educational programs are working in Australian schools" (Wyn et al, 2014, p.5). In addition the new Government agency ACARA was formed to oversee management of NAPLAN tests.

ACARA administered the very first NAPLAN to all Year 3, 5, 7 and 9 students in all Australian and Territory schools in May 2008. Subsequently, individual schools have been required to organise the testing process with their students and participate in it on an annual basis, while ACARA has remained in charge of management of NAPLAN tests to this date. The second test was administered in 2009. The results of both 2008 and 2009 NAPLAN tests were returned to schools, students and parents (as depicted by Figure 1). Test results for students, teachers and schools provided raw results as well as those relative to cohort averages across Australian schools. It is important to note that NAPLAN tests were confidential intended and used for self-assessment and self-improvement. The response from schools was overwhelmingly positive.

The scenario however changed in 2010. Aiming to improve "accountability and transparency", ACARA introduced My School web-portal in Jan 2010, making school performance data collected in 2008 and 2009 publicly available. According to ACARA:

*"My School enables parents, school leaders and their communities, educators and members of wider community to: search for schools in their local areas or from any part of the country; view school-level NAPLAN results; compare statistically similar schools; identify schools that are doing well and share successful strategies"* (ACARA, 2014, p.1.)

At the heart of My School operation is the ongoing process of data collection, processing and dissemination, briefly described as follows and depicted by Figure 2. First of all, data used by My School are collected in several different ways: at the point of a child's enrolment when parents are asked about their own education and occupation, as well as from the Australian Bureau of Statistics (ABS), including for example, ABS census of population and housing data for different areas. Each school is also required to provide financial data, including recurrent income, capital expenditure (presented as a total figure and an average amount per student).

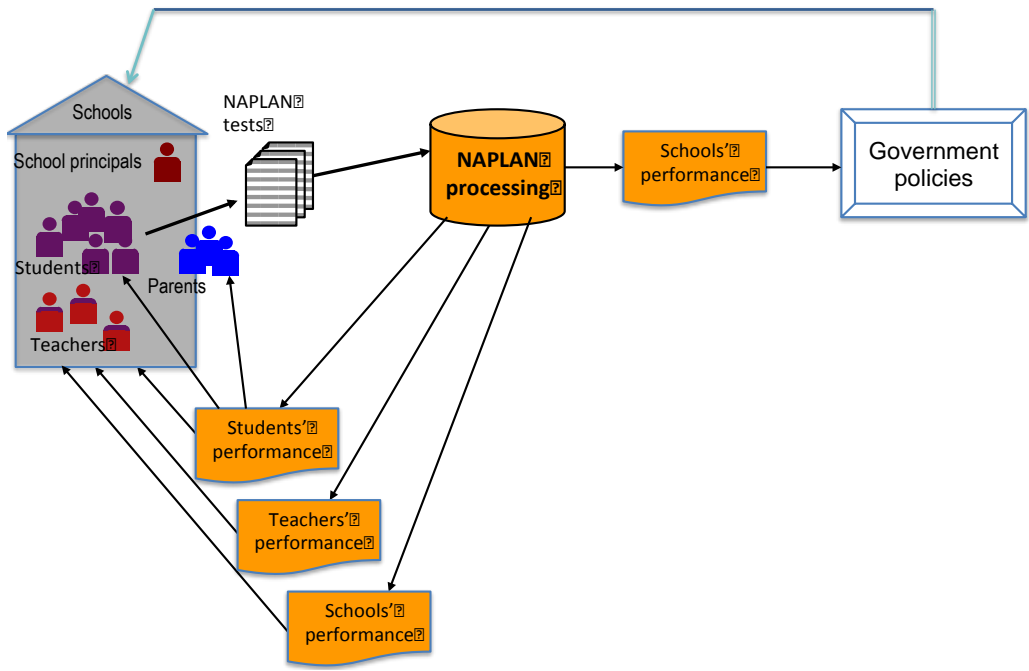

*Figure 1: NAPLAN tests administered from 2008-2010 provided reports to students, teachers and schools about their individual performance*

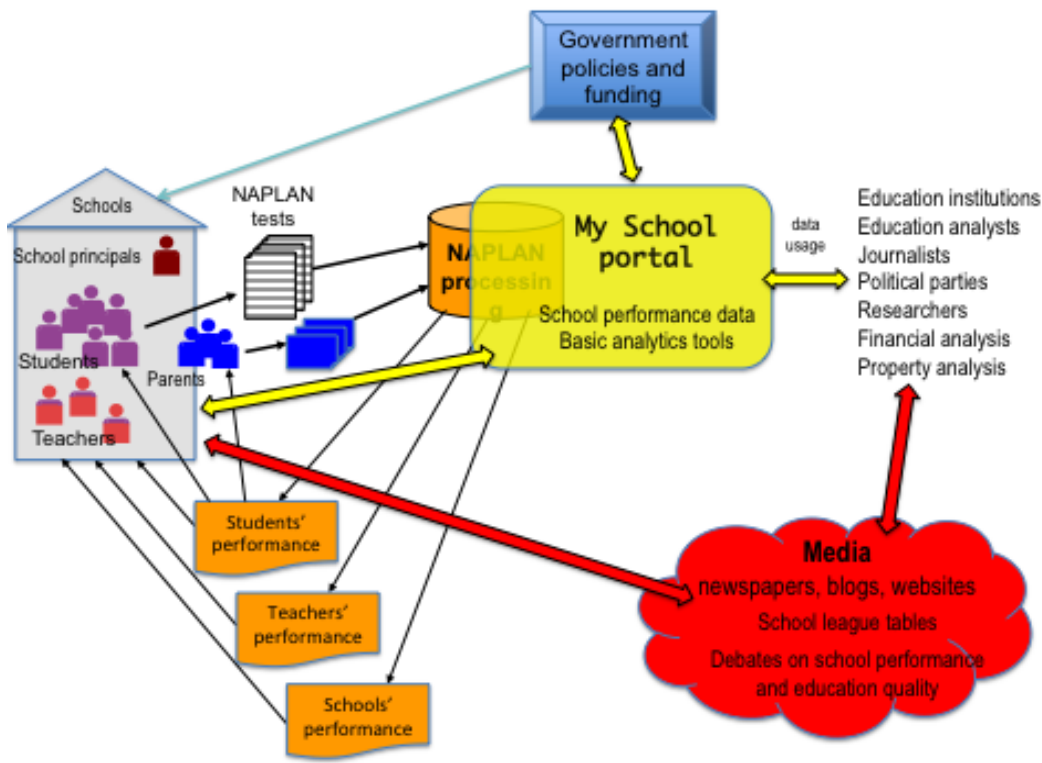

*Figure 2: My School portal launched in Jan 2010 introduced new information flows about schools' performance*

Furthermore, a school profile page includes information about student enrolment, attendance rates as well as staff numbers, which is used to determine school's student/staff ratio. In addition to quantitative data, each school is given an opportunity to provide very brief qualitative information about its values, ethos, programs and main achievements.

However, by far the most important data source for My School comes from NAPLAN tests that are conducted in teacher-supervised classrooms, each year at the same time across Australia. In its current implementation, students' answer booklets are collected and sent to the central Government authority/reporting agency ACARA. This manual process of NAPLAN data collection will be replaced by the so-called "online NAPLAN" from 2017. Following data collection, data is then processed by ACARA, resulting in NAPLAN test results being recorded in a database. In order to enable a more meaningful comparison of students across different schools, as well as grouping of similar schools (of up to 60 schools), ACARA developed an Index of Community Socio-Educational Advantage (ICSEA index), specifically for My School. Development of this index was based on related research showing that family educational and professional background factors (i.e. parents' education level and occupation) are closely related with student educational outcomes. Also to enable meaningful comparison of financial data, ACARA developed a special-purpose financial methodology for My School that takes into account a wide set of variables (e.g. schools' location, type, size, programs and operations).

In the final step, data are then disseminated to individual students, teachers and their schools (within 4-5 months) and soon after school results are made available online on My School. Instead of providing "raw" data as collected from individual schools, data on My School are available in an aggregated (summarized) form at the school level, along with financial, socio-economic data created by ACARA, as described above. All data on My School portal are publicly available and could be analysed using simple tools (also provided).

After My School was launched in Jan 2010 Australian media celebrated the provision of "objective, measurable and reputable data" (Mocker 2013), proclaiming that it will revolutionize education:

*Leaving things up to the experts - keeping performance data secret within the bureaucracy as the critics of publication want - does not result in action. (Sydney Morning Herald Editorial, 2010a)*

*…the test results should be part of wider transparency and accountability about schools and their principals and teachers. The teaching profession should accept that it cannot shield misfits (Sydney Morning Herald Editorial, 2010b).*

*My School, the NAPLAN tests on which it is based and media analysis will revolutionise education by making it possible to base decisions on data not the education establishment's dogma. They establish a marvellous model for other public services from universities to hospitals where consumers have a right to know which service providers are performing … (The Australian Editorial, 2010).*

*It is a health check for the school system and allows schools to check their performance is on track against others, not to reward or punish but so they can use that knowledge to improve (Ferrari, 2014)*

The data and simple analytics tools available via the website have been used by schools leaders and staff, researchers, analysts, journalists and media, to assess schools' and teachers' performance and compare schools in similar areas or socio-economic circumstances (ACARA 2014). Individual data provided to parents show how their child is performing against a national average. However, controversies regarding the nature, use and effects of the data provided via the My School portal emerged soon after its launch.

The key issue was the use of School test results available on My School to compare schools and also to calculate school league tables. In particular newspaper publications of school league tables changed the public discourse on the quality of education and what it means to be a good school. Lower ranked Schools became 'bad' school, students with below average NAPLAN results became 'bad' students and similarly teachers whose students showed below average results became 'bad' teachers.

Numerous negative effects on children, teachers and schools were published and publicly debated (The Advertiser Editorial, 2010; The Australian Editorial, 2010; Anderson, 2010; Sydney Morning Hearld Editorial, 2010a). For example, just within the first few months of this website 'going live' over 480 articles and editorials were published by national and regional newspapers many of which reported negative implications of My School (Mocker, 2013). In addition, numerous submissions to the first Senate inquiry initiated in May 2010 voiced serious concerns:

*A system previously promoting a 'love-of-learning' in a child-centred environment thus sacrificed 'quality education' for 'data-based schooling' (Senate Inquiry 2010, Submission 20).*

*A parent of a child attending a school with below average NAPLAN scores commented on what she saw as 'labelling' students as low achievers, expressing a fear that her daughter and others like her would simply accept the label and stop trying to do better (Senate Inquiry, 2010, Submission 83).*

Similar complaints and individual testimonies of negative implications for students, teachers and pedagogy were reported in the second Senate enquiry in 2013.

What was intended to lead to Australia's 'education revolution' was instead reported to lead to the systematic and comprehensive elimination of everything else previously deemed 'good and valuable' to children's education (Senate Inquiry, 2010, Submission 73). As pointed out by the president of the NSW Teachers Federation "We have never believed that publication of test data out of context means much" (McDougal, 2015). In fact, the most recent reports indicate that "student performances have flatlined in the national literacy and numeracy exams" (Bita, 2015, p.1), or even more seriously, "primary school students have made minimal improvements and high school students have slipped" (Smith, 2015, p.1). These and many other criticisms of NAPLAN and especially of the impacts of public availability of school NAPLAN results on the My School web portal (and school league tables derived from these data) remain unanswered. While these are unintended and unpredicted negative consequences no serious attempts have been made to address them. ACARA continues to claim (supported by many media reports) that My School is highly useful providing equitable benefits to all in the education sector. On each side of the debate considerable evidence and arguments were provided that confuse rather than clarify the issues of social and ethical consequences of My School.

## 5 Theorizing IT-enabled information flows: intermediaries or mediators

To understand how and why different social and ethical consequences show up in the debates on the effects of IT-enabled information flows within the education sector, and what are the implications of these differences, we analyse and interpret the evidence from our My School study. We first examine how My School instigated new information flows in the education sector and then explain how the underlying view on these flows as intermediaries informed one side of the debate on their social and ethical implications leading as it did to an inability to resolve conflicting views and address serious ethical consequences. We then offer an alternative possibility to conceptualize information flows as mediators which enable a different understanding of the effects emerging in the education sector following the introduction of My School. We demonstrate how such understanding helps explain the production of harmful effects on schools, students and teachers, thus enabling a different articulation of ethical questions.

Information flows enabled by NAPLAN processing before the launch of My School in Jan 2010 (presented in Figure 1) included the collection of test results from schools and their processing to produce cohort averages, followed by dissemination of individual results to students and their parents as well as aggregated results to teachers and schools. As mentioned earlier individual test results and comparisons with the cohort average were confidential and provided to each student (and their parents) for self-evaluation. Similarly this was done for teachers and schools so that they could compare their students test results with average school performance across Australia. These information flows can be seen as intermediaries in a sense that they transmitted test results, data (individual and average) from which information could be derived straightforwardly. Individual students (and schools) could see how much they outperformed or underperformed the national average. This type of performance feedback systems has been implemented elsewhere with varying degrees of success (Visscher and Coe 2003; Earl and Katz 2006; Vanderlinde et al. 2010). Apart from the attention to data content and distribution of feedback reports, ethical considerations emphasised confidentiality of data as vital for such systems to achieve their purpose and the desired effects. However as Vanderlinde et al. (2010) warn the production, transmission and use of data have to be continuously monitored to identify potential unintended negative effects.

Information flows radically changed when NAPLAN test results for individual schools became publicly available on the My School web portal (as of Jan 2010). New information flows emerged through the use of this data by a variety of users beyond individual students (and their parent), teachers and schools (as depicted in Figure 2). Importantly publication of school league tables in newspapers triggered a debate on good and bed schools, reasons for poor performance, and necessary measures to address inadequate performance. The debate was dominated by the rhetoric of transparency and responsibility of schools to provide good education. My School was praised for enabling parents to exercise their rights to know the quality of schools and make the "right choices for their children".

The public debate about schools' performance and school league tables resonated loudly in schools, especially at the bottom end of the tables. Many schools experienced serious damage as their reputation was negatively affected by the ranking. Furthermore, underperforming students, labelled 'bed' students, became undesirable and perceived as 'failures' by teachers, parents and themselves (Senate inquiry, 2010). Teachers of underperforming students became 'bad' teachers, some of whom reported repercussions in their schools. Students, teachers and schools took various actions to improve their standing in league tables, thus producing further negative consequences for education and wellbeing of children.

As this short discussion indicates, in the new scenario after the launch of the My School portal (Figure 2) information flows enabled and fostered by My School web portal are not any more efficient means for transmitting school performance data and comparing schools using analytic tools. Information flows have evolved and could not be considered intermediaries any more. They are not simply transporting the data to the broader public as it is widely assumed. Rather information flows instigated by My School are mediators as they are translating the simple NAPLAN test data into particular information about quality of schools, teachers and education. The translation is performed through numerous processes of selection, comparison, aggregation, abstraction, and interpretation involving numerous actors (media, education institutions, political parties, journalists). Due to ensuing public discourse schools, teachers and students were reconstructed as good or bed, desirable or undesirable, with serious consequences for many.

On the other hand according to ACARA the My School portal serves public interest as everybody benefits by getting information about students' and schools' performance:

*My School is a valuable online tool to help educators and communities understand what is happening in schools right across Australia. The site is designed to make it easy for users to access and share information on things such as a school's profile, academic performance, funding sources and financials. You can also see enrolment numbers and attendance rates (ACARA, 2015, p.1).*

ACARA maintains that after the introduction of the My School portal, information flows did not basically change. In other words, it is assumed that information flows remained intermediaries. Consequently My School and the information flows it enabled were justified by the data content being made available. In other words it is claimed that by defining and justifying the inputs (NAPLAN test data) we can define and justify the outputs (information derived from the data by different users). The questions of fairness, justice and ethics, if raised at all, are seen to be related to the data and analytic tools available on My School portal. For instance, the first Senate Inquiry recommended "reforms to the publication and representation of test data, … reforms to the My School web site and management of publications of league tables in media" (Senate Inquiry 2010). Furthermore, Senate recommendation 9 called for examination and public reporting on ways to mitigate the harm caused by simplistic and often distorted information in league tables published by newspapers. In response, ACARA has strengthen legal and technical protections of data and the new version My School 2.0 has new login requirements and terms and conditions to protect the integrity of data (Australian Government, 2011). ACARA will be supported to take steps to counter any inaccurate use of My School information, including pubic response with corrected data.

The problem however is that the use of data publicly available on the My School portal and the enactment of information flows cannot be predicted or controlled. Numerous institutions, analysts, journalists, political parties and the media are all drawing from My School data, interpreting them and deriving 'truthful' information about schools' performance that transform and transcend the original meanings attributed to data at the time of their collection (administration of NAPLAN tests). Processing and interpreting data and presenting reports and information in the media (and by numerous other users) involve a transformative process underpinned by values, interests, political views and sometimes profit motives (e.g. by property analysts). The data are selected, compared, calculated and aggregated by applying certain criteria in order to 'present' and 'reveal' phenomena of interest: the uneven quality of schools, the problem of underperforming schools and their responsibilities, measures to address low quality, and the like. That the NAPLAN data are simple literacy and numeracy test results have been forgotten as the public attention is drawn to the school league tables and comparisons of schools' performance in particular regions and across the country.

As these performance measures travel (through newspapers and other media) what remains hidden is the translation of the narrow and simple literacy and numeracy tests (NAPLAN data) into the measure of schools' quality. In such a way the original intentions and the meaning of the tests are lost (abstracted) and the new meanings ascribed (constructed): NAPLAN test results are translated into the overall measure of quality (disregarding all other possible performance measures) so that schools become high/low quality schools attested by the league tables. Given that a school's 'quality' is calculated based on NAPLAN test results, individual students' results are (re)interpreted as their 'quality' (that is, individual results acquire new meaning within My School's enabled information flows). This suggests that My School's information flows cannot by any stretch of the imagination be understood as intermediaries. They perform as mediators as they "transform, translate, distort, and modify the meaning or the elements they are supposed to carry" (Latour 2005, p. 39). Their outputs or implications cannot be anticipated, predicted or justified based on their inputs.

An important implication of the My School information flows that was neither intended nor predicted is the ongoing reconstruction of the education field. Schools are judged by analysts, commentators, Government funding bodies and parents by their published performance scores and ranking in the league tables, interpreted as objective and legitimate. Schools on their part are trying hard to improve their scores and become 'good' schools (by taking a variety of actions including 'teaching to test' and discouraging low performing students to come to school on the test day). In other words schools are now being constructed according to their image as it is performed by My School information flows, so becoming this image. In an important sense the *map is mistaken for the territory*, to use Alfred Korzybski's language. The map, that is, the image of schools' performance enabled and circulated via the My School portal and more recently via the Australian newspaper portal (that publishes school league tables) is loosing its 'referential being' as it becomes reality itself. In such a way the map, the image "precedes the territory .. [it] engenders the territory" (Baudrillard, 1994, p.1).

The harmful effects of My School (evidenced in the two Senate inquiries and media reports) have not been denied but the public debate and Senate recommendations have not as yet produced more clarity regarding the ethics of My School nor did they succeed in addressing the harmful effects. The dominant underlying view on information flows as intermediaries fostered the debate on the 'causes' of harmful effects (for instance the "wrong use of data") and the responsibilities of users to interpret and use data properly. Such views limited the debate on the ethics of My School's impacts to justification of data content and intended benefits from public availability of data regarding schools performance in NAPLAN test. What remains hidden is that the use and interpretation of the data from My School by the media, numerous analysts and individual journalists, educational institutions and professional associations, schools, teachers, students and their parents, all play important roles in enacting and expanding the information flows. This suggests that these information flows are mediators that translate, reinterpret and distort meanings in unpredicted and uncontrollable ways. The point is not to apportion blame or engage in accusations about one or other of the parties. Rather, as our analysis shows harmful effects result from the ongoing performative co-construction of the My School enabled information flows and practices by multiple actors. Such understanding then allows a different articulation of ethical questions informed by ontological disclosure that reveals and opens to scrutiny mutual reconfiguration of information flows and user's practices as well as their complex effects. The conceptual views on My School and its information flows proposed and discussed in this paper provide a foundation for public exploration of roles and responsibilities, not only by ACARA as an owner of My School portal, but by all actors involved.

## 6 Conclusion

In this paper we explored how different views on public sector IT systems and IT-enabled information flows allow us to see differently their social implications and to articulate the different ethical issues involved (Introna 2007). After defining two opposing views on IT-enabled information flows – conceptualized as 'intermediaries' or 'mediators' (Callon 1991; Latour 1992, 2005) – we engaged them in examining the case of the My School portal. Our findings and analysis demonstrate how the view on My School information flows as intermediaries, the view that underlies the public debate so far, constrains the possibility of revealing and addressing serious social implications and ethical concerns.

Our analysis further demonstrates how understanding My School information flows as mediators radically changes the perspective and the possibility for revealing and considering the questions of fairness and ethics. We show that the assumption that the inputs (data collected and made available via My School portal) determine the outputs (the effects of data usage by the public) is not tenable. Our investigation of harmful effects of My School and how they are performed and enacted through data use, interpretation and circulation via information flows (that are not static but dynamic and expending) demonstrates how the understanding of information flows as mediators alters the ethical perspective on My School. The point is not just to question ACARA's position and the argument that everybody benefits. The key lesson is that the alternative ethical perspective on My School we offer enables ontological disclosure in the complex and dynamic production of a new social order and reconfigurations of the field of education. Furthermore, such ontological disclosure opens to scrutiny numerous practices and the roles and responsibilities of all involved.

The key theoretical contribution of the paper is better understanding of ethical issues arising in IT-enabled information flows in a public sector and how different conceptual views on the information flows allow different articulation and assessment of their ethical implications. The paper also has implications for the debate on My School and similar IT-enabled information flows in public sector, with particular emphasis on responsibilities of all involved (Government agency, media, schools, education institutions, teachers, students and parents).

By way of concluding, we suggest that the paper opens a new theoretical frontier for exploring the questions of fairness and ethics of IT systems and their information flows in society as an important domain for future research. We call for more case studies of other public sector IT systems in health care, education and social services and the ethical challenges created by their information flows, in particular those enabled by open government data. Future work should explore possible approaches for ethically mindful IT-enabled information flows in society seeking "the good life within one's community" (Mingers and Walsham 2010, p. 841).

# Appendix

| Timeline | Significant events | Relevant (selected) documents |
|---|---|---|
| **Initiation of NAPLAN** | | |
| July 2006 | The Ministerial Council on Education, Employment, Training and Youth Affairs (MCEETYA) formally endorsed the introduction of national testing in Literacy and Numeracy | *ACARA Media Releases:*<br>***2008***<br>*- National Curriculum Board Framing Papers Released for Feedback;*<br>*- National Curriculum Consultation Begins Today ;*<br>*- National Curriculum Journey Begins;*<br>***2009***<br>*- ACARA welcomes inaugural Chief Executive Officer;*<br>*- Key features in national curriculum examined at forums;*<br>***ACARA Update Archive****: Issue 1, 27 Nov 2009 and Issue 2, 14 Dec. 2009;* |
| During 2007 | Development of the NAPLAN (National Assessment Program - Literacy and Numeracy) test | |
| May 2008 | ACARA administered the first test | |
| May 2009 | ACARA administered the second test | |
| **Launch of My School online portal** | | |
| Jan 2010 | ACARA makes My School portal available online (with 2008 and 2009 NAPLAN results at the level of individual schools) | *ACARA media Releases, 2010:*<br>*My School website launch;*<br>*Statement from ACARA Chair Professor Barry McGaw;*<br>*National Consultation on the draft Australian Curriculum March 2010;*<br><br>***ACARA Update archive:*** *Issues 3-15 (March – Dec, 2010);* |
| May 2010 | Third NAPLAN test administered, results made available on My School in Sept 2010. | |
| **First Senate Inquiry** | | |
| 13 May 2010 | The Senate referred the matter of NAPLAN to the Senate Education, Employment and Workplace Relations Reference Committee | *268 written submissions to the Senate Inquiry, June 2010;*<br>*Interim report: "Effectiveness of the National Assessment Program – Literacy and Numeracy", Aug. 2011;*<br>*Transcript of the public hearing (84 pages) – Friday 29 Oct 2010 Canberra: Official Committee Hansard Senate: Education, Employment and Workplace Relations References Committee - Reference: National Assessment Program-Literacy and Numeracy;*<br>*Transcript of public hearing (49 pages) – Mon 1 Nov. 2010. Canberra: Official Committee Hansard Senate: Education, Employment and Workplace Relations References Committee – Reference: Primary Schools for the 21st Century program;*<br>*Final report: Education, Employment and Workplace Relations References Committee: Administration and reporting of NAPLAN testing, Nov. 2010;*<br>*Australian Government Response to the Senate Education, Employment and Workplace Relations Reference Committee: Report on the Administration and Reporting of NAPLAN Testing, Aug 2011.* |
| 25 June 2010 | Public submissions open – 268 submissions received | |
| 27 July 2010 | Interim report prepared | |
| 29 Oct & 1 Nov 2010 | Public hearing in Canberra | |
| 24 Nov 2010 | Final report released | |
| May 2011 | Fourth test administered | |
| Aug 2011 | Australian Government responds | |
| **My School post 1st Senate Inquiry** | | |
| May 2012 | Fifth suite of tests administered | ***ACARA media releases, 2013:***<br>*SCSEEC (from 1 July 2014 known as Education Council) media release – 2013 NAPLAN National Report*<br>　　*2013 NAPLAN Summary Report release*<br>　　*Delay in release of 2013 NAPLAN Student Reports*<br><br>***ACARA Update archive for 2012****– Issues 36-58 (Feb – Dec, 2010);* ***ACARA Update archive for 2013****– 12 updates (Feb – Dec, 2010)* |
| May 2013 | Sixth suite of tests administered | |
| **Second Senate Inquiry** | | |
| 15 May 2013 | The Senate referred the matter of NAPLAN to the Senate Education, Employment and Workplace Relations References | *93 written public submissions to the Senate Inquiry -June 2013;*<br>*Interim report:* ***The effectiveness of the National Assessment Program Literacy and Numeracy*** |
| | Committee for inquiry and report. | *(NAPLAN) – 27 June 2013*<br>*Transcript of the public hearing (53 pages) – 21 June 2013*<br>*Melbourne: Official Committee Hansard Senate: Education, Employment and Workplace Relations References Committee - Reference: National Assessment Program-Literacy and Numeracy;*<br>*Final report: The Senate: Education and Employment Reference Committee: Effectiveness of the National Assessment Program – Literacy and Numeracy: Final Report, March 2014;*<br>*Australian Government Response to the Senate Education, Employment and Workplace Relations Reference Committee: Report on the Effectiveness of NAPLAN Testing, Jun 2014;* |
| **7 June 2013** | **Public submissions close** | |
| 21 June 2014 | Public hearing in Melbourne | |
| **27 June 2013** | **Interim report: The effectiveness of the National Assessment Program – Literacy and** Numeracy (NAPLAN) | |
| 27 March 2014 | Final report released | |
| Jun 2014 | Australian Government responds | |
| **My School post-2nd Senate inquiry** | | |
| May 2014 | Seventh suite of tests administered (results made available on My School in Sept 2014) | *ACARA Media releases:*<br>*Fair comparisons: My School website released for 2015;*<br>    *NAPLAN 2015: the last paper-based tests for some*<br>    *National Assessment and Surveys Online Program: tailored test design 2013 study;*<br>    *NAPLAN summary information released;*<br>    *NAPLAN tests start tomorrow;*<br>    *ACARA releases statement to the review of Australian Curriculum;*<br>    *Release of My School 2014;*<br>**ACARA Update archive for 2014**– 27 update documents (Feb – Dec, 2014);<br>**ACARA Update archive for 2015**– 12 update documents (Jan – June 2015); |
| March 2015 | Plans announced to introduce online testing from 2017 | |
| 22 March 2015 | Australian Government Review of My School announced (still in progress) | |
| (expected) 2017 | Online testing | |

*Table 2: Significant events and relevant documents in My School case study*

## Acknowledgments

The authors would like to thank Dr. Patricia Morgan, Post-Doctoral Research Fellow (UNSW and University of Canberra) for her useful comments and suggestions.

## Copyright

**Copyright:** © 2015 Cecez-Kecmanovic, D. and Marjanovic, O. This is an open-access article distributed under the terms of the Creative Commons Attribution-NonCommercial 3.0 Australia License, which permits non-commercial use, distribution, and reproduction in any medium, provided the original author and ACIS are credited.